\documentclass[a4paper,11pt]{article}
\usepackage{jheppub} 
\usepackage{lineno}

\title{\boldmath Why Cooper pairs live in AdS$_{2}$: a spectral analysis of the Yukawa--SYK
model }

\author{Veronika C. Stangier$^{(a)}$ and}
\affiliation{$^{(a)}$ Institute for Theory of Condensed Matter, Karlsruhe Institute of Technology,
Karlsruhe 76131, Germany}
\author{J{\"o}rg Schmalian$^{(a),(b)}$}
\affiliation{$^{(b)}$ Institute for Quantum Materials and Technologies, Karlsruhe Institute
of Technology, Karlsruhe 76131, Germany}

\emailAdd{joerg.schmalian@kit.edu}

\abstract{
We establish the spectral foundation of the geometric formulation
of Cooper pairing in the Yukawa--Sachdev--Ye--Kitaev model. Starting
from the large-$N$ bilocal effective action, we derive the Gaussian
fluctuation kernel in the even-frequency spin-singlet Cooper channel.
Following  the construction of Maldacena
and Stanford,  we determine the kernel eigenvalue $k\left(h\right)$
analytically for arbitrary conformal weight $h$ and resolve the pairing
fluctuations into the continuous and discrete sectors of the associated
de-Sitter space Laplacian. We show that the superconducting instability and the universal low-energy fluctuations near it reside entirely in the continuous scattering sector, while the discrete modes remain non-critical. Expanding the kernel about the lowest continuum mode yields a Klein--Gordon action on ${\rm dS}_2$, restricted to the continuous spectral subspace.
This is precisely the subspace on which the inverse Radon transform to ${\rm AdS}_2$ is well defined. The projected bilocal Cooper-pair field can therefore be mapped onto a scalar matter field propagating in ${\rm AdS}_2$, without requiring any further spectral restriction on the bulk theory. Our results provide the missing microscopic justification for the projection implicit in earlier holographic formulations and clarify how a local bulk field emerges from the bilocal pairing theory.}

\begin{document}
\maketitle
\flushbottom

\section{Introduction}

The Sachdev--Ye--Kitaev (SYK) model has emerged as a paradigmatic
example of a strongly interacting quantum many-body system without
quasiparticle excitations~\cite{Sachdev1993,Georges2000,Georges2001,Sachdev2010,Kitaev2015,Kitaev2015b,Sachdev2015,Maldacena2016,Chowdhury2022}. Its combination of random all-to-all interactions of $N$ fermion species,
analytical tractability in the large-$N$ limit, emergent conformal
symmetry, and maximal quantum chaos has made it an important point
of contact between condensed-matter physics, quantum information,
and quantum gravity.
At energies well below the microscopic interaction scale, the fermionic
two-point function approaches a scale-invariant saddle point, while the dominant corrections are governed by soft reparametrization modes, whose dynamics is described by a Schwarzian action. Remarkably, the
same Schwarzian theory arises as the boundary dynamics of nearly anti-de
Sitter (AdS) space in Jackiw-Teitelboim (JT) gravity~\cite{Jackiw1985,Teitelboim1983,MaldacenaStanfordYang2016,Jensen2016,Engelsoy2016}.
In this low-energy regime, the SYK model and two-dimensional JT gravity
provide two descriptions of the same universal infrared dynamics.
The SYK model therefore offers a setting in which central ideas of
holography can be studied starting from an explicitly defined and
microscopically controlled quantum-mechanical Hamiltonian.

For applications to quantum critical matter, a useful extension is
provided by the Yukawa-SYK (Y-SYK) model~\cite{Esterlis2019,Wang2020,Hauck2020,Classen2021}.
Here, $N$ fermionic degrees of freedom are coupled to $M$ dynamical
bosons through random Yukawa interactions. In the appropriate large-$N$
and large-$M$ limit, with the ratio $M/N$  held fixed, the saddle-point equations form a closed set
of Schwinger-Dyson equations. In the pairing channel, these equations
take the form of strongly coupled Eliashberg equations that become asymptotically exact.
The normal state is an incoherent non-Fermi liquid with scale-invariant fermionic and bosonic correlations, in which the singular retarded interaction mediated by the critical bosons can induce Cooper pairing and superconductivity -- of course in the sense of a mean-field theory with all-to-all interactions. The model thus allows for studying superconductivity in the absence
of long-lived electronic quasiparticles and, more broadly, for clarifying
the relation between quantum-critical Eliashberg theory and holographic
descriptions of superconducting instabilities. In addition, it has
been extended to finite dimensional systems and used to analyze the
interplay of quantum criticality, quantum chaos, and superconductivity
in a range of systems~\cite{Esterlis2021,Tikhanovskaya2022,Patel2023,Li2024,Guo2024,Kim2021,StangierSheehy2026,Kim2026,StangierTwist2026,Esterlis2026}.

A direct relation between the many-body field theory and holographic
perspectives of the superconducting degrees of freedom was discussed
in Refs.~\cite{Inkof2022,StangierFS2026}. Starting from the effective
action for Gaussian fluctuations of the bilocal Cooper-pair field,
it was shown that the low-energy pairing problem can be mapped onto
a local Gaussian field theory in two-dimensional Euclidean  anti-de Sitter space~\footnote{In this paper, we consider exclusively  Euclidean ${\rm AdS}_2$ and, for simplicity, will not explicitly indicate its  Euclidean character each time. By contrast, ${\rm dS}_2$ always refers to the usual (Lorentzian) de Sitter space, even though we work in imaginary time.}.
In this construction, the anomalous pair field
\begin{equation}
F\left(\tau_{1},\tau_{2}\right)=-\frac{1}{N}\sum_{i=1}^Nc_{i\uparrow}\left(\tau_{1}\right)c_{i\downarrow}\left(\tau_{2}\right)
\end{equation}
is directly related  to a field 
\begin{equation}
    \Psi\left(\tau_{1},\tau_{2}\right)\propto\left|\tau_{1}-\tau_{2}\right|^{2\Delta}F\left(\tau_{1},\tau_{2}\right)
    \label{eq:Psi_first0}
\end{equation}
on the space of pairs of time variables, where $\Delta$ is the fermion scaling dimension.  The center-of-mass time plays the role of the boundary time,
whereas the relative time---or, equivalently, its conjugate relative
frequency---encodes the emergent radial direction.
According to Refs.~\cite{Inkof2022,StangierFS2026}, the Gaussian pairing theory of the Y-SYK model and the matter sector of an ${\rm AdS}_{2}$ holographic superconductor are not merely analogous: within the low-energy approximation, they are related by an explicit integral transformation. The main reasoning  is that an effective low-energy action 
\begin{equation}
S_{{\rm pair}}\left[\Psi\right]= N\int\sqrt{-g_{{\rm dS}_2}}d^2\zeta \,\Psi^\dagger \left(\tilde{m}^{2}-\Box_{{\rm dS}_{2}}\right)\Psi
\label{eq:act_kin}
\end{equation}
of the  scalar field  $\Psi$ of Eq.~\eqref{eq:Psi_first0} in de Sitter space ${\rm dS}_{2}$ can be derived from the Yukawa-SYK theory. 
Here, $\zeta$ is a suitable choice of ${\rm dS}_{2}$ coordinates. One then applies a Radon transform~\cite{Helgason2011,Das2018,Stone2025}
$\Psi={\cal R}\psi$ to express $\Psi$ in terms of a field $\psi$ in  ${\rm AdS}_{2}$. The integral Radon transform ${\cal R}$,  averages the field over geodesics in  ${\rm AdS}_2$; see Appendix~\ref{app:Radon}. An important property of the Radon transform is that it intertwines the two Laplacians $\square_{{\rm dS}_{2}}{\cal R}\psi={\cal R}\left(\square_{{\rm AdS}_{2}}\psi\right)$. 
This property allows one to demonstrate that the action $S_{{\rm pair}}$ can be expressed in terms of that of a Gaussian field in  anti-de Sitter space:
\begin{equation}
S_{\rm pair}\left[\Psi\right]\propto {\cal S}_{\rm pair}\left[\psi\right],
\label{eq:action_to_action}
\end{equation}
with
\begin{equation}
{\cal S}_{{\rm pair}}\left[\psi\right]= N\int \sqrt{g_{{\rm AdS}_2}}d^2\xi\, \psi^\dagger \left(m^{2}-\Box_{{\rm AdS}_{2}}\right)\psi
\label{eq:SH4}
\end{equation}
and $\xi$ coordinates in ${\rm AdS}_2$. 
In Eq.~\eqref{eq:SH4}, the mass $m$ differs from $\tilde{m}$ of Eq.~\eqref{eq:act_kin}, although the two agree at the Breitenlohner-Freedman bound~\cite{breitenlohner1982positive}. Clearly, constructing the field in ${\rm AdS}_2$ requires the Radon transform to be invertible. For convenience, the details of this holographic map are summarized in Appendix~\ref{app:map}.

The resulting correspondence therefore implies that the scale-invariant
normal state determines the emergent ${\rm AdS}_{2}$ background,
while the Cooper-pair field becomes a matter field living on this
geometry. 
Contributions to the pairing kernel arising from the incoherent
fermionic particle-particle propagator and from the singular boson-mediated
attraction combine into the effective bulk mass $m$ of the scalar field.
The onset of superconductivity is mapped onto reaching the ${\rm AdS}_{2}$
Breitenlohner-Freedman stability bound. Pair breaking tunes the
effective mass through this bound and thereby produces the characteristic
quantum phase transition of the Y-SYK superconductor. At the Gaussian level, the matter field does not back-react on the gravitational sector. However, finite temperature
introduces an ${\rm AdS}_{2}$ black-hole horizon, deviations from
particle-hole symmetry generate a background electric field, and sources
coupled to the microscopic Cooper pair determine boundary conditions
for the bulk scalar~\cite{Inkof2022}. This may be viewed as a step towards a microscopic understanding of holographic superconductivity as proposed in Refs.~\cite{Gubser2008,hartnoll2008,hartnoll2008b}.
%

There is, however, a subtle restriction implicit in this construction. The inverse Radon transform, and hence our ability to relate  generic field configurations $\Psi \leftrightarrow  \psi$ between  ${\rm dS}_{2}$ and ${\rm AdS}_{2}$,   is not defined for an arbitrary function on ${\rm dS}_{2}$~\cite{,Das2018,Stone2025}. 
This issue becomes transparent upon diagonalizing the ${\rm dS}_{2}$ Laplacian. The eigenfunctions of the (negative) Laplacian in ${\rm dS}_{2}$, defined by $-\Box_{{\rm dS}_2} \Psi_\nu = \mu_\nu \Psi_\nu$, form both continuous and discrete parts of the spectrum with eigenvalues
\begin{eqnarray}
\mu(p)&=&\frac{1}{4}+p^{2}, \hspace{3.1cm} p>0 \nonumber \\
\mu_{n}&=&-\left(2+2n\right)\left(1+2n\right), \hspace{1cm} n\in  \mathbb{N}.
\label{eq:eigenvals}
\end{eqnarray}
The  eigenvalue problem  in ${\rm dS}_2$ can be recast as an inverted Liouville quantum-mechanics problem. Here, the continuous states correspond to scattering states and the discrete ones to bound states, whose energy depends on the precise self-adjoint extension chosen to regularize the singular potential~\cite{Andrianov2018,Stone2025}. As we discuss in Appendix~\ref{app:Radon}, the Radon transform is only invertible in the continuous part of the ${\rm dS}_2$-spectrum. Indeed, the spectrum of $-\Box_{{\rm AdS}_2} \psi_\nu = \mu_\nu \psi_\nu$ in  ${\rm AdS}_2$ consists only of the continuous part. Functions that generate the discrete spectrum upon Radon transform are not properly normalizable in ${\rm AdS}_2$~\cite{Stone2025}. In the holographic construction of Refs.~\cite{Inkof2022,StangierFS2026}, the discrete part of the ${\rm dS}_{2}$ spectrum was therefore effectively removed ``by hand''.  However, if one replaces $-\Box_{{\rm dS}_2} $ in Eq.~\eqref{eq:act_kin} by its discrete eigenvalues $\mu_{n}$, the resulting theory is clearly unstable. Removing the discrete spectrum amounts to ignoring what appears to be an infinite number of highly unstable modes.
While this restriction gives reasonable results,  its formal justification within the Y-SYK pairing problem has so far remained incomplete. 
\begin{figure}
    \centering
    \includegraphics[width=0.85\linewidth]{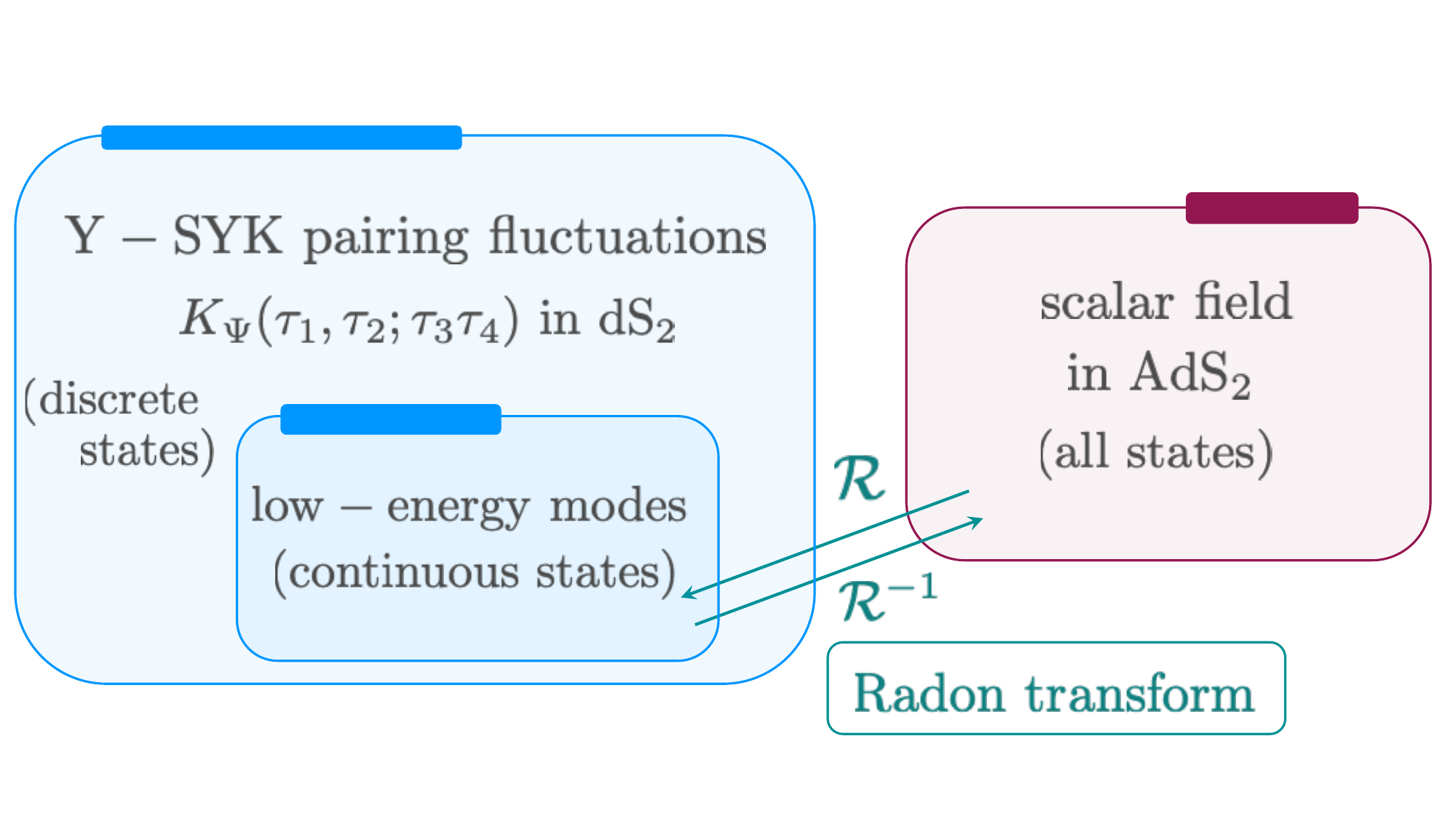}
    \caption{Schematic illustration of the holographic map between pairing fluctuations in the Y-SYK model and a scalar field in ${\rm AdS}_{2}$. Pairing fluctuations of  scale-invariant fermions, governed by the pairing kernel $K_{\Psi}(\tau_{1},\tau_{2};\tau_{3},\tau_{4})$, span the two-dimensional de Sitter space ${\rm dS}_{2}$. This space contains both continuous and discrete eigenstates of $K_\Psi$. The discrete states remain non-critical w.r.t. the onset of superconductivity and can therefore be excluded from the universal low-energy theory of fluctuating Cooper-pairs. The remaining continuous sector can be mapped, via an invertible Radon transform $\mathcal{R}$, onto the full Euclidean anti-de Sitter space ${\rm AdS}_{2}$, which thus provides the natural geometric formulation of the critical low-energy theory of the pairing sector.
}
    \label{fig:schem}
\end{figure}

The purpose of the present work is to close this gap. We analyze the eigenfunctions of the conformal Casimir and the pairing kernel in the full bilocal function space, keeping track of both the continuous and discrete parts of the spectrum. We demonstrate that the physical low-energy continuum theory of pairing fluctuations occupies only a specific subspace of the  Casimir Hilbert space. In particular, the discrete modes are shown  not to contribute to the infrared Cooper-pair field. The microscopic Y-SYK theory therefore selects precisely the sector that admits a geometric transformation to a local scalar theory in ${\rm AdS}_{2}$; see Fig.~\ref{fig:schem} for a schematic illustration. This provides the missing formal justification for the restriction implicit in the construction of Refs.~\cite{Inkof2022,StangierFS2026}. It also shows that ${\rm AdS}_{2}$, unlike ${\rm dS}_{2}$,  does not require an additional projection onto any spectral subspace. Cooper pairing in quantum-critical systems can therefore be formulated most naturally in anti-de Sitter space.

Our proof builds on the conformal-kernel
formalism developed by Maldacena and Stanford for the SYK four-point
function~\cite{Maldacena2016}. Their approach exploits the commutation
of the conformal ladder kernel with the generators of ${\rm SL}(2,\mathbb{R})$,
allowing the kernel to be diagonalized in a basis of Casimir eigenfunctions
and the four-point function to be resolved into continuous and discrete
spectral contributions. We adapt this machinery to the Cooper channel
of the Y-SYK model. In doing so, we identify  which Casimir eigenfunctions are
actually relevant  in the low-energy regime of the pairing problem. It follows that the discrete spectrum does not contribute to the critical part of the theory and can  be projected out. This
analysis not only places the previously proposed ${\rm AdS}_{2}$
formulation on firmer mathematical ground, but also clarifies more
generally how the physical field content of an emergent holographic
theory is selected from the larger space of bilocal collective fields of the Yukawa-SYK theory.

\section{The Yukawa-SYK model}

We consider $N$ flavors of spin-$1/2$ fermions $c_{i,\sigma}$ ($i=1\cdots N$, $\sigma=\pm$) coupled to $M$ real
bosons $\phi_l$ ($l=1\cdots M$) through random Yukawa matrix elements
\begin{equation}
H_{\rm Y-SYK}=-\mu\sum_{i,\sigma}c^\dagger_{i\sigma}c_{i \sigma}+
\frac{1}{N}\sum_{ijl,\sigma}g_{ij,l}\phi_{l}c_{i\sigma}^{\dagger}c_{j\sigma}+\frac{1}{2}\sum_{l}\left(\pi_l^2+\omega_0^2\phi_l^2\right).
\end{equation}
$\pi_l$ is the momentum conjugate to $\phi_l$. $\omega_0$ is the bare boson frequency and $\mu$ the chemical potential. In this paper we consider $\mu=0$. 
The random couplings $g_{ij,l}$ have zero mean and a variance chosen such that
the interaction survives in the large-$N$ and large-$M$ limit with $M/N$
fixed. Following Ref.~\cite{Hauck2020}, we use complex-valued couplings
$g_{ij,l}={\rm Re} g_{ij,l}+i {\rm Im}g_{ij,l}$ with mean square $\overline{\left({\rm Re}g_{ij,l}\right)^{2}}=\left(1-\frac{\alpha}{2}\right)g^{2}$
and $\overline{\left({\rm Im}g_{ij,l}\right)^{2}}=\frac{\alpha}{2}g^{2}$, with $0\leq \alpha \leq 1$. In the limit $\alpha=0$, the coupling constants are real and, for each given $l$, distributed according to a Gaussian orthogonal ensemble. By contrast, $\alpha=1$ corresponds to complex couplings distributed according to a Gaussian unitary ensemble. Hence, $\alpha >0$ corresponds to a state where,
for a given realization of the $g_{ij,l}$, time-reversal symmetry is absent, and only
restored on average. Therefore,  $\alpha$ serves as a tool
to continuously enhance pair breaking; see Ref.~\cite{Hauck2020} for
a detailed discussion of the phase diagram as a function of $\alpha$. In short, the ground state is superconducting for small $\alpha$, while superconductivity disappears above  a critical pair breaking strength, which for $M=N$ is $\alpha_c\approx 0.6265$.

After disorder averaging, one introduces bilocal collective fields
for the fermionic and bosonic propagators. This is most efficiently
done using non-redundant Nambu spinors 
\begin{equation}
c_{i}=\left(c_{i\uparrow},c_{i\downarrow}^{\dagger}\right)^{T}.
\end{equation}
The bilocal Nambu fields  are
\begin{equation}
\hat{G}\left(\tau_{1},\tau_{2}\right)=-\frac{1}{N}\sum_{i=1}^{N}c_{i}\left(\tau_{1}\right)c_{i}^{\dagger}\left(\tau_{2}\right)=\left(\begin{array}{cc}
G\left(\tau_{1},\tau_{2}\right) & F\left(\tau_{1},\tau_{2}\right)\\
F^{\dagger}\left(\tau_{1},\tau_{2}\right) & \tilde{G}\left(\tau_{1},\tau_{2}\right)
\end{array}\right),
\end{equation}
where $\tilde{A}\left(\tau,\tau'\right)=-A\left(\tau',\tau\right).$
The conjugate bilocal field is the Nambu self-energy
\begin{equation}
\hat{\Sigma}\left(\tau_{1},\tau_{2}\right)=\left(\begin{array}{cc}
\Sigma\left(\tau_{1},\tau_{2}\right) & \Phi\left(\tau_{1},\tau_{2}\right)\\
\Phi^{\dagger}\left(\tau_{1},\tau_{2}\right) & \tilde{\Sigma}\left(\tau_{1},\tau_{2}\right)
\end{array}\right).
\end{equation}
For the bosons one introduces in full analogy 
\begin{equation}
D\left(\tau_{1},\tau_{2}\right)=-\frac{1}{M}\sum_{l=1}^{M}\phi_{l}\left(\tau_{1}\right)\phi_{l}\left(\tau_{2}\right)
\end{equation}
and its conjugate  $\Pi\left(\tau_{1},\tau_{2}\right)$. 
The effective action after replica-symmetric averaging over the random
couplings is
\begin{eqnarray}
S & = & -N{\rm Tr}\log\left(\hat{G}_{0}^{-1}-\hat{\Sigma}\right)+\frac{M}{2}{\rm Tr}\log\left(D_{0}^{-1}-\Pi\right)\nonumber \\
 & - & N{\rm Tr}\left(\hat{\Sigma}\otimes\hat{G}\right)+\frac{M}{2}{\rm Tr}\left(\Pi\otimes D\right)+S_{{\rm int}},
 \label{eq:full_action}
\end{eqnarray}
where $\left(A\otimes B\right)(\tau_1,\tau_3)=\int d\tau_{2}A\left(\tau_{1},\tau_{2}\right)B\left(\tau_{2},\tau_{3}\right)$.
$\hat{G}_{0}^{-1}\left(\tau_{1},\tau_{2}\right)=-\partial$$_{\tau_{1}}\delta\left(\tau_{1}-\tau_{2}\right)\hat{1}$
and $D_{0}^{-1}\left(\tau_{1},\tau_{2}\right)=\left(-\partial_{\tau_{1}}^{2}+\omega_{0}^{2}\right)\delta\left(\tau_{1}-\tau_{2}\right)$
are the two bare propagators, respectively. 
For the interaction term, disorder averaging produces 
\begin{equation}
S_{{\rm int}}=Mg^{2}{\rm Tr}\left[(G\tilde{G})\otimes D-\left(1-\alpha\right)\left(F^{\dagger}F\right)\otimes D\right] .\label{eq:Sint}
\end{equation}

\begin{figure}
    \centering
    \includegraphics[width=0.75\linewidth]{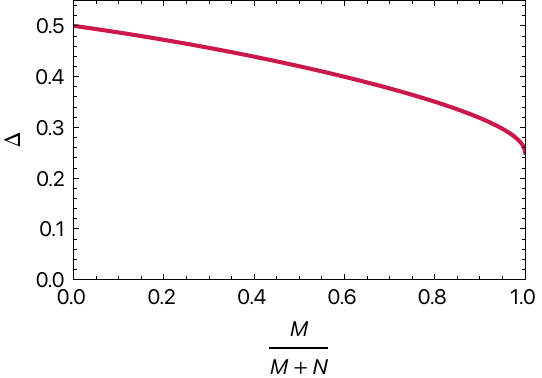}
    \caption{Exponent $\Delta$ for varying ratio of the boson and fermion flavor numbers $M$ and $N$, respectively, where $\Delta(M\ll N)\rightarrow 1/2-M/(8N)$ and $\Delta(M\gg N)\rightarrow 1/4+\left(N/(8\pi M)\right)^{1/2}$.}
    \label{fig:Delta}
\end{figure}

The analysis of the saddle point that results from this action was
performed in Refs.~\cite{Esterlis2019,Wang2020,Hauck2020,Classen2021}. In the normal state
with $F=\Phi=0$ it follows that at long times 
\begin{equation}
G\left(\tau\right)  =  -A_{G}\frac{{\rm sign}\left(\tau\right)}{\left|\tau\right|^{2\Delta}} \qquad
D\left(\tau\right)  =  A_{D}\frac{1}{\left|\tau\right|^{2-4\Delta}},
\label{eq:propagators}
\end{equation}
with coefficients $A_{G}=\frac{\Gamma\left(2\Delta\right)\sin\left(\pi\Delta\right)}{\pi A_{\Sigma}}$
and $A_{D}=\frac{A_{\Sigma}^{2}}{g^{2}}\frac{\pi\left(1-4\Delta\right)\cos2\pi\Delta}{8\Gamma\left(2\Delta\right)^{2}\cos\left(\pi\Delta\right)\sin^{3}\left(\pi\Delta\right)}.$ Although the bosons are initially massive, a critical massless state emerges at low energies for all values of the coupling constant $g$, albeit with a coupling-dependent extent of the critical regime~\cite{Esterlis2019,Wang2020}.
The scaling exponent $\Delta$ of the fermions is  fixed by the condition
\begin{equation}
\frac{4\Delta-1}{2\left(2\Delta-1\right)\left(\sec\left(2\pi\Delta\right)-1\right)}=\frac{N}{M}.\label{eq:condDelta}
\end{equation}

For $M=N$, for example, one obtains $\Delta\approx 0.42037$. The variation of $\Delta$ between $1/4$ and $1/2$ for different values of the ratio $M/N$ is shown in Fig.~\ref{fig:Delta}.
$A_{\Sigma}$ is the coefficient of the self energy in frequency space
$\Sigma\left(\omega\right)=-iA_{\Sigma}{\rm sign}\left(\omega\right)\left|\omega\right|^{1-2\Delta}$
and cannot be determined within the low-energy scaling theory, but follows from a numerical solution of the  saddle-point equations. Here, we will not need a result for $A_{\Sigma}$ as it cancels
in the important combination
\begin{equation}
\kappa\equiv g_{p}^{2}A_{G}^{2}A_{D}=\frac{\left(1-\alpha\right)\left(1-2\Delta\right)}{2\pi}\tan\left(\pi\Delta\right),
\end{equation}
where we also used the relation  Eq.~\eqref{eq:condDelta} for the exponent $\Delta$.
$g_{p}^{2}=g^{2}\tfrac{M}{N}\left(1-\alpha\right)$ is the interaction in
the pairing channel; see Eq.~\eqref{eq:Sint}. 

\section{Gaussian action of Cooper pairs}

Next we focus on Gaussian fluctuations in the pairing fields on top
of the scale invariant normal state described by Eq.~\eqref{eq:propagators}. If we expand the action $S$ of Eq.~\eqref{eq:full_action} to second order in $F$ and $\Phi$ and their conjugates, it follows $S\approx S_0+ S_{\rm pair}$, where $S_0$ contains the saddle point and potential fluctuations of $G$ and $\Sigma$. The Gaussian pairing term is
 \begin{eqnarray}
S_{{\rm pair}} & = & N{\rm Tr}\left(\tilde{G}\otimes\Phi^{\dagger}\otimes G \otimes \Phi-g_{p}^{2}F^{\dagger}F\otimes D-F^{\dagger}\otimes\Phi-F\otimes\Phi^{\dagger}\right).
\label{eq:firstGaussianaction}
\end{eqnarray}
The saddle-point equations derived from this expression are
\begin{eqnarray}
F\left(\tau_{1},\tau_{2}\right) & = & -\int_{3,4}G\left(\tau_{2}-\tau_{4}\right)G\left(\tau_{1}-\tau_{3}\right)\Phi\left(\tau_{3},\tau_{4}\right),\nonumber \\
\Phi\left(\tau_{1},\tau_{2}\right) & = & -g_{p}^{2}F\left(\tau_{1},\tau_{2}\right)D\left(\tau_{1}-\tau_{2}\right),
\end{eqnarray}
where $\int_{3,4}\cdots$ is shorthand for $\int d\tau_3\int d\tau_4\cdots $. It yields the linearized ``gap'' equation of the superconductor:
\begin{equation}
\Phi\left(\tau_{1},\tau_{2}\right)=\int_{3,4}K_{\Phi}\left(\tau_{1},\tau_{2};\tau_{3},\tau_{4}\right)\Phi\left(\tau_{3},\tau_{4}\right),
\label{eq:lgaptime}
\end{equation}
where the pairing kernel $K_\Phi$ is given by a combination of the critical  fermion  and   boson propagators:
\begin{eqnarray}
K_{\Phi}\left(\tau_{1},\tau_{2};\tau_{3},\tau_{4}\right)&=&g_{p}^{2}G\left(\tau_{2}-\tau_{4}\right)G\left(\tau_{1}-\tau_{3}\right)D\left(\tau_{1}-\tau_{2}\right).
\end{eqnarray}
If we use the power-law solutions, Eq.~\eqref{eq:propagators}, for $G$ and $D$, we
obtain
\begin{equation}
K_{\Phi}\left(\tau_{1},\tau_{2};\tau_{3},\tau_{4}\right)=\frac{\kappa\,{\rm sign}\left(\tau_{2}-\tau_{4}\right){\rm sign}\left(\tau_{1}-\tau_{3}\right)}{\left|\tau_{2}-\tau_{4}\right|^{2\Delta}\left|\tau_{1}-\tau_{3}\right|^{2\Delta}\left|\tau_{1}-\tau_{2}\right|^{2-4\Delta}}.
\end{equation}

For our later discussion, it will be useful to also analyze the linearized
equation for the anomalous propagator
\begin{equation}
F\left(\tau_{1},\tau_{2}\right)=\int_{3,4}K_{F}\left(\tau_{1},\tau_{2};\tau_{3},\tau_{4}\right)F\left(\tau_{3},\tau_{4}\right)
\end{equation}
with the slightly different kernel
\begin{eqnarray}
K_{F}\left(\tau_{1},\tau_{2};\tau_{3},\tau_{4}\right)&=&g_{p}^{2}G\left(\tau_{2}-\tau_{4}\right)G\left(\tau_{1}-\tau_{3}\right)D\left(\tau_{3}-\tau_{4}\right)\nonumber \\
    & = & K_{\Phi}\left(\tau_{1},\tau_{2};\tau_{3},\tau_{4}\right)\left|\frac{\tau_{1}-\tau_{2}}{\tau_{3}-\tau_{4}}\right|^{2-4\Delta}.\label{eq:KF}
\end{eqnarray}
Regardless of which equation we want to solve, a pairing instability  corresponds to the largest eigenvalue of either $K_\Phi$ or $K_F$ reaching unity.

For the development of a Gaussian field theory, it is convenient   to introduce yet another bilocal field:
\begin{eqnarray}
\Theta\left(\tau_{1},\tau_{2}\right) & = & g_{p}\sqrt{D\left(\tau_{1}-\tau_{2}\right)}F\left(\tau_{1},\tau_{2}\right),\nonumber \\
\Theta^{\dagger}\left(\tau_{1},\tau_{2}\right) & = & g_{p}\sqrt{D\left(\tau_{1}-\tau_{2}\right)}F^{\dagger}\left(\tau_{1},\tau_{2}\right).
\label{eq:Theta_field}
\end{eqnarray}
Integrating out the $\Phi$ field and expressing $F$ in terms of $\Theta$, we find that the action is given by
\begin{eqnarray}
S_{{\rm {\rm pair}}} & = & N\int d\tau_{1}\cdots d\tau_{4}\Theta^{\dagger}\left(\tau_{1},\tau_{2}\right)\left(\delta_{\tau_{1}-\tau_{3}}\delta_{\tau_{2}-\tau_{4}}-K_{\Theta}\left(\tau_{1},\tau_{2};\tau_{3},\tau_{4}\right)\right)\Theta\left(\tau_{3},\tau_{4}\right)\nonumber \\
 & = & N\Theta^{\dagger}\left(1-K_{\Theta}\right)\Theta.
 \label{eq:act_pair_full}
\end{eqnarray}
In the last line we used an obvious  operator notation where the rows and columns of $K_{\Theta}$ are labeled by the index pair $(\tau_1,\tau_2)$  and $(\tau_3,\tau_4)$, respectively. 
The explicit form of  $K_\Theta$ is 
\begin{eqnarray}
K_{\Theta}\left(\tau_{1},\tau_{2};\tau_{3},\tau_{4}\right)&=&g_{p}^{2}\sqrt{D\left(\tau_{1}-\tau_{2}\right)}G\left(\tau_{2}-\tau_{4}\right)G\left(\tau_{1}-\tau_{3}\right)\sqrt{D\left(\tau_{3}-\tau_{4}\right)}, \nonumber \\
& = & K_{\Phi}\left(\tau_{1},\tau_{2};\tau_{3},\tau_{4}\right)\left|\frac{\tau_{1}-\tau_{2}}{\tau_{3}-\tau_{4}}\right|^{1-2\Delta}.  
\label{eq:KThe}
\end{eqnarray}
Similar to the above cases, an instability occurs when an eigenvalue  of $K_{\Theta}$ reaches
$1$, while the normal state  is stable against pairing fluctuations if all eigenvalues of $K_\Theta$ are smaller than $1$. 

\section{Spectrum of the pairing kernels}
The three kernels are related by similarity transformations and hence share eigenvalues:
\begin{equation}
    K_{\Phi}\Phi_\nu=k_\nu\Phi_\nu,\qquad K_{F}F_\nu=k_\nu F_\nu, \qquad K_{\Theta}\Theta_\nu=k_\nu\Theta_\nu.
    \label{eq:all_kernels}
\end{equation}
 To see this we introduce the similarity transform of a function $J\left(\tau_{1},\tau_{2}\right)$
\begin{equation}
\left(M_{a}J\right)\left(\tau_{1},\tau_{2}\right)=\left|\tau_{1}-\tau_{2}\right|^{a}J\left(\tau_{1},\tau_{2}\right),
\label{eq:similarity}
\end{equation}
which in the index-pair notation is merely the multiplication
of $J$ with a diagonal matrix. It then holds that the operators are also related by similarity transforms, 
\begin{eqnarray}
K_{F} & = & M_{2-4\Delta}K_{\Phi}M_{2-4\Delta}^{-1},\nonumber \\
K_{\Theta} & = & M_{1-2\Delta}K_{\Phi}M_{1-2\Delta}^{-1},
\end{eqnarray}
i.e. they all share the same eigenvalues. 
One can of course draw the same conclusion using the explicit expressions Eqs.~\eqref{eq:KF} and \eqref{eq:Theta_field} and finds that the eigenfunctions are
simply related via 
\begin{eqnarray}
\Theta_{\nu}\left(\tau_{1},\tau_{2}\right) & = & \left|\tau_{1}-\tau_{2}\right|^{1-2\Delta}\Phi_{\nu}\left(\tau_{1},\tau_{2}\right),\nonumber \\
F_{\nu}\left(\tau_{1},\tau_{2}\right) & = & \left|\tau_{1}-\tau_{2}\right|^{2-4\Delta}\Phi_{\nu}\left(\tau_{1},\tau_{2}\right).
\label{eq:eigenf_rel}
\end{eqnarray}

In order to determine the eigenvalues $k_\nu$ we exploit the  connection between the eigenfunctions of the above
kernels and the eigenfunctions of the Laplacian in de Sitter space
$d{\rm S}_{2}$, following the approach by Maldacena and
Stanford in Ref.~\cite{Maldacena2016}. In fact, Ref.~\cite{Maldacena2016} analyzed exactly the kernel $K_{F}$ introduced in Eq.~\eqref{eq:KF}, a connection that was also pointed out in Ref.~\cite{Khveshchenko2026}. The  difference to their analysis is merely that  Ref.~\cite{Maldacena2016} was interested in eigenfunctions that are  odd under the exchange of the two time arguments while we are interested in  an even-frequency pairing state, i.e. $F(\tau_1,\tau_2)=F(\tau_2,\tau_1)$. Most of the steps  carry over naturally. First, Ref.~\cite{Maldacena2016} showed that $K_{F}$ commutes with the Casimir
operator 
\begin{equation}
C_{1+2}=2\left(\Delta^{2}-\Delta\right)-J_{1}P_{2}-P_{1}J_{2}+2D_{1}D_{2}
\end{equation}
 of the $\mathfrak{sl}(2,\mathbb{R})$ algebra with $D_{i}=-\tau_{i}\partial_{\tau_{i}}-\Delta$,
$P_{i}=\partial_{\tau_{i}}$, and $J_{i}=\tau_{i}^{2}\partial_{\tau_{i}}+2\Delta\tau_{i}$. The statement
$\left[C_{1+2},K_{F}\right]=0$ implies that the two operators share eigenfunctions. 
Let the eigenvalues and eigenfunctions
of the Casimir be given as $C_{1+2}F_{\nu}=-\mu_{\nu}F_{\nu}$, where it
is convenient to express the eigenvalues as $\mu_\nu=-h_\nu\left(h_\nu-1\right)$. 
Moreover, $K_F$ commutes with the total generators of $\mathfrak{sl}(2,\mathbb R)$ acting on the bilocal field,
i.e. $\left[K_{F},D_{1}+D_{2}\right]=\left[K_{F},P_{1}+P_{2}\right]=\left[K_{F},J_{1}+J_{2}\right]=0$.
Schur’s lemma then implies that, within each irreducible representation, the eigenvalue of $K_F$ depends only on the conformal weight $h$, i.e. there exists a relation $k_{\nu}=k\left(h_\nu\right)$~\cite{Maldacena2016}. Our goal is to determine this relation for our problem. 

To make the connection to de Sitter space explicit, we introduce the center-of-mass and relative coordinates
\begin{equation}
   t  =  \frac{\tau_{1}+\tau_{2}}{2} \qquad z  =  \frac{\tau_{1}-\tau_{2}}{2},
   \label{eq:dScoordinates}
\end{equation}
such that $\tau_{1,2}=t\pm z$. If one then considers a function
\begin{equation}
    F\left(t,z\right)=F_0\left|z\right|^{-2\Delta}\Psi\left(t,z\right),
    \label{eq:Psi_first}
\end{equation}
with arbitrary coefficient $F_0$, it holds that
\begin{equation}
C_{1+2}F\left(t,z\right)=F_0\left|z\right|^{-2\Delta}z^{2}\left(\partial_{z}^{2}-\partial_{t}^{2}\right)\Psi\left(t,z\right).
\end{equation}
With the ${\rm dS}_2$ metric
\begin{equation}
ds^{2}=\frac{dz^{2}-dt^{2}}{z^{2}}
\end{equation}
follows for the Laplacian $\square_{{\rm dS}_{2}}=z^{2}\left(\partial_{z}^{2}-\partial_{t}^{2}\right)$.
Hence, we can express the ${\rm dS}_2$-Laplacian   in terms of the  Casimir via another similarity transform
\begin{equation}
    \square_{{\rm dS}_{2}}=M_{2\Delta}C_{1+2}M_{2\Delta}^{-1}.
\end{equation}
This implies $\square_{{\rm dS}_{2}}\Psi_{\nu}=-\mu_{\nu}\Psi_{\nu}=h_\nu\left(h_\nu-1\right)\Psi_{\nu} $, i.e. the Laplacian and the Casimir share 
the same eigenvalues.

Let us now consider the functions 
\begin{eqnarray}
F_{h,\tau_{0}}\left(\tau_{1},\tau_{2}\right) & = & \frac{{\rm sign}\left(\tau_{1}-\tau_{0}\right){\rm sign}\left(\tau_{2}-\tau_{0}\right)\left|\tau_0\right|^{2h}}{\left|\tau_{1}-\tau_{0}\right|^{h}\left|\tau_{2}-\tau_{0}\right|^{h}\left|\tau_{1}-\tau_{2}\right|^{2\Delta-h}}.
\label{eq:eigenf_gen}
\end{eqnarray}
The  difference to the functions used in Ref.~\cite{Maldacena2016} is, one the one hand, that the anomalous Gor'kov function $F_{h,\tau_{0}}$ does not carry an additional factor ${\rm sign}(\tau_1-\tau_2)$,  since we consider even-frequency pairing. On the other hand, we included the factor $\prod_{i=1,2}{\rm sign}\left(\tau_{i}-\tau_{0}\right)$ to ensure that $F_{h,\tau_{0}}\left(\tau_{1},\tau_{2}\right)$ is an eigenfunction.
One easily finds that   the $F_{h,\tau_{0}}$ are eigenfunctions of the Casimir 
\begin{equation}
C_{1+2}F_{h,\tau_{0}}=h\left(h-1\right)F_{h,\tau_{0}}.
\end{equation}
The proof that $F_{h,\tau_{0}}$ is also an eigenfunction of the full kernel is given in Appendix~\ref{app:eigenfct_proof}.
Here, $\tau_{0}$ plays the role of a continuous quantum number, similar
to the frequency that follows after Fourier transformation w.r.t.
the center-of-mass time $t$. The eigenvalues only
depend on the exponent $h$, which labels the irreducible representations of ${\rm SL}(2,\mathbb{R})$. Hence,  $\nu=\{h,\tau_0\}$ determine the quantum numbers of the eigenstates, which are degenerate w.r.t. $\tau_0$. The generalization of Ref.~\cite{Maldacena2016} to our problem of exchange-symmetric functions is straightforward since the Casimir and the kernel $K_{F}$
both preserve the exchange symmetry. Hence
we can simply apply  $K_F$ to $F_{h,\tau_0}$ and determine the dependence $k\left(h\right)$ of the eigenvalues. 

\begin{figure}
    \centering
    \includegraphics[width=0.75\linewidth]{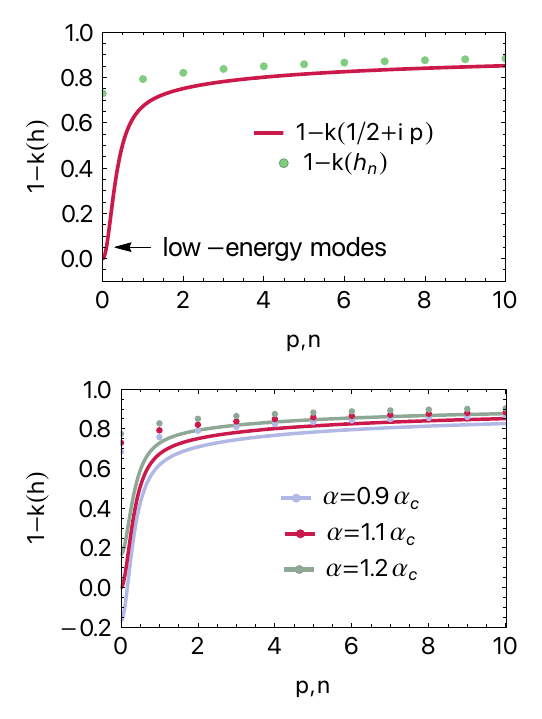}
    \caption{ $1-k\left(h\right)$, with $k(h)$ the eigenvalue of the 
  the mutually similar pairing kernels $K_F$, $K_\Phi$, $K_\Theta$ or $K_\Psi$, as function of $p$ (for the
continuous spectrum with $h=\frac{1}{2}+ip$)  and as function
of $n$ (for the discrete spectrum with $h=h_{n}=2\left(n+1\right)$).
Upper panel: The continuous (discrete) eigenvalues are shown in red (green). We find that  critical  pairing fluctuations are only due to the continuous part of the spectrum. Discrete modes are noncritical.
Results are shown  for $\alpha=\alpha_{c}$ where the largest
eigenvalue $k\left(h\right)$ reaches $1$.
Lower panel: continuous and discrete eigenvalues for different values of the pair-breaking strength $\alpha$.  Notice that, for $\alpha=0.9\alpha_c$, the ground state is superconducting and the largest eigenvalue obeys $k(h)>1$, i.e. the theory is already unstable.}
    \label{fig:EV_as_f_of_EV}
\end{figure}

Owing to the degeneracy of the Casimir, we may evaluate $K_{F}F_{h,\tau_{0}}$ for a convenient choice of $\tau_0$.
Following Ref.~\cite{Maldacena2016} we  send $\tau_{0}\rightarrow\infty$ to find
\begin{eqnarray}
F_{h,\infty}\left(\tau_{1},\tau_{2}\right) & = & \frac{1}{\left|\tau_{1}-\tau_{2}\right|^{2\Delta-h}}.
\label{eq:f_trial}
\end{eqnarray}
Using the kernel $K_{F}$ of Eq.~\eqref{eq:KF}, we analyze $K_{F}F_{h,\infty}=k(h)F_{h,\infty}$ and obtain
for  the eigenvalues:
\begin{eqnarray}
k\left(h\right) & = & \kappa\int d\tau d\tau'\frac{{\rm sign}\left(1-\tau\right){\rm sign}\left(-\tau'\right)}{\left|1-\tau\right|^{2\Delta}\left|\tau'\right|^{2\Delta}\left|\tau-\tau'\right|^{2-2\Delta-h}}.
\label{eq:kofh_integral}
\end{eqnarray}
The evaluation of the integral is  straightforward and yields 
\begin{equation}
    k(h)=\frac{\left(1-\alpha\right)\pi\Gamma\left(2-2\Delta\right)}{\Gamma\left(2-h-2\Delta\right)\Gamma\left(1+h-2\Delta\right)\Gamma\left(2\Delta\right)\left(\sin\left(h\pi\right)-\sin\left(2\pi\Delta\right)\right)}.
    \label{eq:eigenvals_K_fin}
\end{equation}
With this result, we can express the kernel $K_{F}$ as a function of the Casimir operator
\begin{equation}
 K_{F}=k\left(\frac{1}{2}+\sqrt{\frac{1}{4}+C_{1+2}}\right).   
 \label{eq:operator_fct}
\end{equation}
Notice that the  integral Eq.~\eqref{eq:kofh_integral} converges only for $1-2\Delta<{\rm Re}(h)<2\Delta$, but the result Eq.~\eqref{eq:eigenvals_K_fin} can be analytically continued for the  discrete spectrum.
Using  $\mu=-h(h-1)$ for the eigenvalues, we can generate the continuous and discrete parts of the spectrum of Eq.~\eqref{eq:eigenvals} with
\begin{eqnarray}
    h&=&\frac{1}{2}+ ip,\qquad p>0 \nonumber \\
    h_{n}&=&2(n+1),\qquad n\in\mathbb{N}.
    \label{eq:h_values}
\end{eqnarray}
Let us  analyze the function $k(h)$ of Eq.~\eqref{eq:eigenvals_K_fin}. In Fig.~\ref{fig:EV_as_f_of_EV}, we show $1-k\left(h\right)$ for $h=\frac{1}{2}+ip$ (the continuous states) and $h=h_{n}$ (the discrete states).  Most notably, if compared to the  Laplacian $-\square_{{\rm dS}_2}$, 
the operator $1-K_F$ is now bounded from below and the discrete and continuous eigenvalues reverse their order.
Furthermore, long-wavelength theory such as Eq.~\eqref{eq:act_kin} follows from the complete Gaussian theory of Eq.~\eqref{eq:act_pair_full} only if the expansion captures the lowest eigenmodes of $1-K_F$. Fig.~\ref{fig:EV_as_f_of_EV} shows that the critical low-energy sector is confined to the space spanned by the continuous states. Notice, even though their eigenvalues should not themselves be interpreted as physical excitation energies, the discrete modes are high-energy modes of the Gaussian pairing theory in the sense that their inverse susceptibilities $1-k(h_n)$ remain finite at the transition. The separation of the Cooper-pair excitation spectrum into a critical continuous sector and a non-critical discrete sector, shown in Fig.~\ref{fig:EV_as_f_of_EV}, is the central result of this paper.

\begin{figure}
    \centering
    \includegraphics[width=0.95\linewidth]{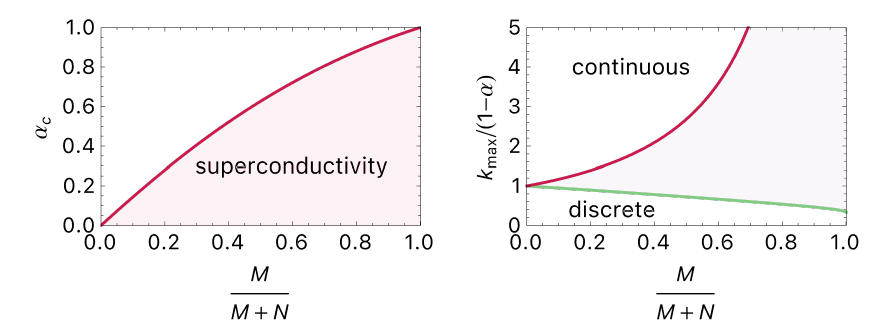}
    \caption{Left Panel: Critical value $\alpha_c$ of the pair-breaking parameter $\alpha$ for varying ratio of the boson and fermion flavor numbers $M$ and $N$, respectively. Right panel: maximal eigenvalue  of the  continuous part of the spectrum $k\left(1/2\right)$ (red), corresponding to $p=0$ in $h=\frac{1}{2}+ip$, and of the discrete part of the spectrum  $k\left(2\right)$  (green), corresponding to $n=0$ in $h=2(n+1)$. Unless $M\ll N$ and superconductivity is extremely fragile against pair breaking, the two parts of the spectrum are well separated for small $p$. The eigenvalues are divided by $1-\alpha$ for clarity.}
    \label{fig:alphac}
\end{figure}

The condition $k(1/2)=1$, corresponding to $p=0$, determines the instability of the lowest mode. It follows 
\begin{equation}
k_0\equiv k(1/2)=\frac{
\pi(1-\alpha)\Gamma\left(2-2\Delta\right)
}{
\Gamma^{2}\left(\frac{3}{2}-2\Delta\right)
\Gamma\left(2\Delta\right)
\left[1-\sin\left(2\pi\Delta\right)\right]
}.
\label{eq:k0expr}
\end{equation}
This is in full agreement with the results of Refs.~\cite{Esterlis2019,Hauck2020}. For example, inserting $\Delta\approx 0.42037 $ yields the critical pair-breaking strength $\alpha_c\approx 0.6265$ mentioned above from the condition $k_0=1$. The $M/N$-dependence of $\alpha_c$ is shown in the left panel of Fig.~\ref{fig:alphac}. For $\alpha<\alpha_c$ the system is in the superconducting phase, which is destroyed by pair breaking at $\alpha=\alpha_c$. Increasing the number of boson modes stabilizes superconductivity. In the right panel of Fig.~\ref{fig:alphac} we show the largest eigenvalue $k_0$ of the continuous part of the spectrum in comparison with the largest discrete eigenvalue $k(2)=(1-\alpha)\frac{\Delta}{1-\Delta}$. Unless $M\ll N$, when superconductivity is exceptionally fragile against pair breaking, the two parts of the spectrum remain well separated at small $p$. 

\section{Low energy theory in ${\rm dS}_2$}

Finally, we demonstrate that our analysis indeed generates an effective
field theory in de Sitter space. We recall the Gaussian action for pairing fluctuations
given in Eq.~\eqref{eq:act_pair_full} in terms of the field $\Theta$. We express $\Theta$ in
terms of $\Psi$ of Eq.~\eqref{eq:Psi_first0}. 
Both $\Theta$ and $\Psi$ are related to $F$ through similarity transformations and are therefore related to each other according to
\begin{equation}
\Theta=M_{-1}\Psi,
\end{equation}
where $M_{a}$ is defined in Eq.~\eqref{eq:similarity}. Expressed in terms of the two time variables, this relation is simply
$\Theta\left(\tau_{1},\tau_{2}\right)=\left|\tau_{1}-\tau_{2}\right|^{-1}\Psi\left(\tau_{1},\tau_{2}\right)$.
The similarity transformation allows us to rewrite the action as
\begin{equation}
S_{{\rm pair}}=N\Psi^{\dagger}M_{-1}^{2}\left(1-K_{\Psi}\right)\Psi,
\label{eq:act_Psi_sim_dd}
\end{equation}
where $K_{\Psi}=M_{-1}^{-1}K_{\Theta}M_{-1}$ has the same eigenvalues
$k\left(h_{\nu}\right)$. We have further used that $M_{a}^{\dagger}=M_{a}$
for a real diagonal matrix.  The explicit expression for $K_\Psi$ is
\begin{eqnarray}
K_{\Psi}\left(\tau_{1},\tau_{2};\tau_{3},\tau_{4}\right)&=&K_{\Phi}\left(\tau_{1},\tau_{2};\tau_{3},\tau_{4}\right)\left|\frac{\tau_{1}-\tau_{2}}{\tau_{3}-\tau_{4}}\right|^{-2\Delta}.  
\label{eq:KPsi2}
\end{eqnarray}

 In Eq.~\eqref{eq:Psi_first} it was shown that $\Psi$ is an element of the space of functions spanned by the eigenstates of the ${\rm dS}_{2}$ Laplacian, i.e. $K_{\Psi}$ acts on the same
 functions as $\square_{{\rm dS_{2}}}$. Therefore,  we obtain in
analogy to Eq.~\eqref{eq:operator_fct}:
\begin{equation}
K_{\Psi} =k\left(\frac{1}{2}+\sqrt{\frac{1}{4}+\square_{{\rm dS}_2}}\right).
\end{equation}
Let us insert this result into Eq.~\eqref{eq:act_Psi_sim_dd} and expand $\Psi=\sum_{\nu}A_{\nu}\Psi_{\nu}$ and, similarly, $\Psi^{\dagger}$ in
terms of the eigenfunctions of $\square_{{\rm dS_{2}}}$. This yields
\begin{equation}
S_{{\rm pair}}=N\sum_{\nu'\nu}A_{\nu'}^{*}A_{\nu}\left(1-k\left(h_{\nu}\right)\right)\Psi_{\nu'}^{\dagger}M_{-1}^{2}\Psi_{\nu}.
\end{equation}
Writing the double-index expression explicitly as an integral over the two time variables gives
\begin{equation}
\Psi_{\nu'}^{\dagger}M_{-1}^{2}\Psi_{\nu}=\int\frac{d\tau_{1}d\tau_{2}}{\left(\tau_{1}-\tau_{2}\right)^{2}}\Psi_{\nu'}^{*}\left(\tau_{1},\tau_{2}\right)\Psi_{\nu}\left(\tau_{1},\tau_{2}\right).
\end{equation}
We observe that this is precisely the integration measure of ${\rm dS}_{2}$ when expressed in
the coordinates $t$ and $z$ introduced in Eq.~\eqref{eq:dScoordinates}. In other words,
in generic ${\rm dS}_{2}$ coordinates we can write
\begin{equation}
\Psi_{\nu'}^{\dagger}M_{-1}^{2}\Psi_{\nu}=\frac{1}{2}\int\sqrt{-g_{{\rm dS}_{2}}}d^{2}\zeta\Psi_{\nu'}^{*}\left(\zeta\right)\Psi_{\nu}\left(\zeta\right)=\frac{1}{2}\delta_{\nu,\nu'},
\end{equation}
where we have used the fact that the eigenfunctions of the Laplacian are orthogonal
with respect to this measure. The factor $\tfrac{1}{2}$ is due to the Jacobian of Eq.~\eqref{eq:dScoordinates}. It then follows that the action takes the form
\begin{eqnarray}
S_{{\rm pair}} & = & \frac{N}{2}\sum_{\nu}A_{\nu}^{*}A_{\nu}\left(1-k\left(h_{\nu}\right)\right)\nonumber \\
& = & \frac{N}{2}\int\sqrt{-g_{{\rm dS}_{2}}}d^{2}\zeta\, \Psi^{\dagger }\left(1-k\left(\frac{1}{2}+\sqrt{\frac{1}{4}+\square_{{\rm dS}_2}}\right)\right)\Psi,
\label{eq:full_reformulation}
\end{eqnarray}
demonstrating that the theory naturally admits a formulation as a field theory in de Sitter space. Up to now, this theory is fully equivalent to our original Gaussian action of Eq.~\eqref{eq:firstGaussianaction} and describes all pairing fluctuations of the scale-invariant critical Y-SYK theory. 

To derive  from Eq.~\eqref{eq:full_reformulation}  a low-energy theory, relevant near the onset of superconductivity, we use our result for the spectrum $k(h)$ and focus on its
continuous part. We therefore project out the non-critical discrete modes and expand the continuous eigenvalue near $p=0$
\begin{equation}
k\left(\frac{1}{2}+ip\right)\approx k_{0}\left(1-b_{0}p^{2}\right).
\end{equation}
Here, $k_{0}$, given in Eq.~\eqref{eq:k0expr}, and $b_{0}=-
\psi^{(1)}\!\left(3/2-2\Delta\right)
+\frac{\pi^{2}}{2-2\sin\left(2\pi\Delta\right)}$ are positive constants~\footnote{$\psi^{(1)}$ is the first derivative of the digamma function}.
 The resulting effective action is
\begin{eqnarray}
S_{\rm pair} & = & \tilde{c}_0 N\int\sqrt{-g_{{\rm dS}_{2}}}d^{2}\zeta\,\Psi_{\rm c}^{\dagger}\left(\tilde{m}^{2}-\square_{{\rm dS}_{2}}\right)\Psi_{\rm c},
\label{eq:act_fin}
\end{eqnarray}
where $\Psi_{\rm c}=P_c\Psi$ denotes the projection onto the continuous spectral subspace. 
The proportionality constant is $\tilde{c}_0=k_{0}b_0/2$, while the mass is
\begin{equation}
\tilde{m}^{2}=\frac{1-k_{0}}{k_{0}b_0}-\frac{1}{4}.
\end{equation}
It reaches the Breitenlohner-Freedman bound $m^{2}_{\rm BF}=-1/4$~\cite{breitenlohner1982positive} precisely at the transition, where the largest eigenvalue $k_0$ reaches $1$.
 In the Introduction, we mentioned that replacing the Laplacian in Eq.~\eqref{eq:act_kin} by the discrete eigenvalues $-\mu_n$ of Eq.~\eqref{eq:eigenvals} leads to an infinite number of unstable modes. It is now clear how this instability is avoided. One must first restrict the field configurations to the space spanned by the continuous states and only then perform an expansion in $\frac{1}{4}+\square_{{\rm dS}_2}$. The aforementioned instability arises only if one first performs the expansion and subsequently allows for arbitrary field configurations.

We have shown that the long-wavelength expansion of the Yukawa-SYK theory yields the action given in Eq.~\eqref{eq:act_kin}, subject to the additional restriction that it be projected onto the continuous sector of ${\rm dS}_2$, as required for a proper justification of the holographic map developed in Refs.~\cite{Inkof2022,StangierFS2026}. The main steps of this map are summarized in Appendix~\ref{app:map}.

\section{Analysis in frequency space}
In order to make contact with the literature on Eliashberg theory,  we discuss some of our conclusions in frequency space.
The eigenfunction Eq.~\eqref{eq:f_trial} corresponds  with Eq.~\eqref{eq:eigenf_rel} to
\begin{eqnarray}
\Phi_{h,\infty}\left(\tau_{1},\tau_{2}\right) & = & \frac{1}{\left|\tau_{1}-\tau_{2}\right|^{2-2\Delta-h}}.
\label{eq:phi_trial}
\end{eqnarray}
For functions  $\Phi\left(\tau_{1},\tau_{2}\right)=\Phi\left(\tau_{1}-\tau_{2}\right)$
follows from Eq.~\eqref{eq:lgaptime} upon Fourier transformation 
\begin{eqnarray}
\Phi\left(\omega\right) & = & g_{p}^{2}\int\frac{d\omega'}{2\pi}D\left(\omega-\omega'\right)G\left(-\omega'\right)G\left(\omega'\right)\Phi\left(\omega'\right).
\end{eqnarray}
With the transformed propagators in frequency space
\begin{equation}
    G\left(\omega\right) =  -\frac{i}{A_{\Sigma}}\frac{{\rm sign}\left(\omega\right)}{\left|\omega\right|^{1-2\Delta}} \qquad D\left(\omega\right)  =  \frac{A_{\Sigma}^{2}}{g^{2}C_{\Delta}}\frac{1}{\left|\omega\right|^{4\Delta-1}}
\end{equation}
follows
\begin{equation}
\Phi\left(\omega\right)=\frac{1-\alpha}{C_{\Delta}}\int\frac{d\omega'}{2\pi}\frac{\Phi\left(\omega'\right)}{\left|\omega-\omega'\right|^{4\Delta-1}\left|\omega'\right|^{2-4\Delta}}
\end{equation}
with $
C_{\Delta}=-8\cos\left(\pi\Delta\right)\sin^{3}\left(\pi\Delta\right)\Gamma\left(2\Delta\right)\Gamma\left(1-2\Delta\right)/\pi^{2}$. This equation has been discussed in great detail in Refs.~\cite{abanov2001,abanov2001b,Chubukov2005,abanov2020-I,abanov2020-II,Wang2013,Wang2015,Wu2019,Wang_Chub_2025} with numerous applications for quantum-critical pairing in compressible and incompressible systems in finite dimensions. 
The eigenfunction Eq.~\eqref{eq:phi_trial} then corresponds for the continuous part of the spectrum and with $h=\frac{1}{2}\pm ip$ to 
\begin{eqnarray}
\Phi\left(\omega\right)&\propto&\frac{1}{\left|\omega\right|^{\frac{4\Delta-1}{2}\pm ip}},
\label{eq:phi_trial_omega}
\end{eqnarray}
solutions that have already been discussed in  Refs.~\cite{abanov2001,abanov2001b,Chubukov2005,abanov2020-I,abanov2020-II,Wang2013,Wang2015,Wu2019,Wang_Chub_2025}. 
Within  Euclidean theory, $1-k(\frac12+ip)$ determines the inverse susceptibility of the corresponding mode, i.e. they enter the partition sum as low-lying excited states.

Using Eq.~\eqref{eq:eigenf_gen}, that describes the entire family of degenerate solutions,  we can construct the corresponding solutions of the gap equation in frequency space, parametrized by $\tau_{0}$. Clearly these solutions do not solely depend on $\tau_1-\tau_2$. Fourier transforming
$\Phi\left(\tau_{1},\tau_{2}\right)$ where $\omega$ is the frequency
associated with $\tau_{1}-\tau_{2}$ while $\Omega$ is the frequency  conjugate to the center-of-mass time $\left(\tau_{1}+\tau_{2}\right)/2$, 
we then obtain a class of degenerate "dynamic" solutions of the linearized
gap equation 
\begin{equation}
\Phi\left(\omega\right)\rightarrow\Phi\left(\omega,\Omega\right)=\Xi\left(\frac{\Omega}{\omega}\right)\frac{e^{i\Omega\tau_{0}}}{\left|\omega\right|^{\frac{4\Delta-1}{2}\pm ip}}
\label{eq:phi_trial_omega2}
\end{equation}
with
\begin{equation}
\Xi\left(y\right)=c\int_{-\infty}^{\infty}\frac{dx}{2\pi}\frac{{\rm sign}\left(\frac{y^{2}}{4}-x^{2}\right)}{\left|\frac{y^{2}}{4}-x^{2}\right|^{1-h}\left|{\rm 1}-x\right|^{2\Delta+h-1}}
\label{eq:actual_frequency_solution}
\end{equation}
and $c=-8\left|\tau_{0}\right|^{2h}\Gamma\left(1-h\right)^{2}\Gamma\left(h+2\Delta-1\right)\cos^{2}\frac{\pi h}{2}\sin\left(\frac{\pi}{2}\left(h+2\Delta\right)\right)$
as well as $h=\frac{1}{2}\mp i p$.  The existence of this degenerate family suggests potentially rich dynamics in the time-dependent pairing response, such as the AC Josephson effect or AC-modulated junction transmissions.

\section{Conclusion}

In this work, we have clarified the spectral origin of the emergent
$\mathrm{AdS}_{2}$ description of Cooper pairing in the Yukawa--SYK model.
Starting from the large-$N$ bilocal effective action, we derived the Gaussian
action for even-frequency spin-singlet pairing fluctuations about the
scale-invariant normal state. The resulting pairing kernels acting on the
anomalous propagator, the anomalous self-energy, and a symmetrized pairing
field are related by similarity transformations and therefore possess the same
eigenvalue spectrum. A superconducting instability occurs when the largest of
these eigenvalues reaches unity.

The central observation underlying our analysis is that the pairing kernel can be related to the Laplacian in de Sitter space.  Using the construction of Maldacena and Stanford in the exchange-even
sector, relevant for even-frequency pairing, we were  able to
diagonalize the kernel in a basis of Casimir eigenfunctions. We  obtained the kernel
eigenvalue as an explicit function of the conformal weight $h$, see Eq.~\eqref{eq:eigenvals_K_fin},
which allows us to treat the continuous and discrete sectors of the 
spectrum on equal footing. Since the Casimir
is equivalent, after a field redefinition, to the scalar Laplacian on
$\mathrm{dS}_{2}$, this also provides a spectral decomposition of the bilocal
Cooper-pair field in the associated kinematic space. 
Hence, one can express the full Gaussian pairing theory as an action of a scalar field in de Sitter space; see Eq.~\eqref{eq:full_reformulation}.

Our main result is that the superconducting instability is controlled entirely
by the continuous scattering sector, $h=\frac{1}{2}\pm ip$. The largest
eigenvalue occurs at the bottom of this continuum, $p=0$, and reaches unity at
the critical pair-breaking strength. Analyzing the full pairing kernel, the discrete states, by contrast,
remain separated from the critical sector as long as the ratio $M/N$ of boson and fermion modes remains finite. They are therefore high-energy modes (non-critical modes with a finite pairing susceptibility)  from the viewpoint
of the universal theory near the superconducting transition and may
consistently be integrated out. This resolves the apparent instability that
would result from applying a naive local $\mathrm{dS}_{2}$ action to the
complete spectrum of the Laplacian. 

Projecting onto the continuous spectral subspace and expanding the kernel
around $p=0$, we obtained the local low-energy action, Eq.~\eqref{eq:act_fin}, in terms of the field $\Psi_c$ that is not an arbitrary function on
$\mathrm{dS}_{2}$, but only the part of the original bilocal field that belongs
to the principal continuous series. This is precisely the range on which the
inverse Radon transform used in the geometric formulation is well defined.
The resulting field can consequently be mapped to a scalar matter field in
 $\mathrm{AdS}_{2}$. Our spectral analysis thus supplies the missing
microscopic justification for the restriction implicit in the holographic
construction of Refs.~\cite{Inkof2022,StangierFS2026}.

The frequency-space formulation provides an equivalent and familiar
interpretation of this result. The continuum eigenfunctions correspond to 
power-law solutions, given in Eq.~\eqref{eq:phi_trial_omega}.
More generally, the complete family of continuum eigenfunctions yields solutions that also depend on the center-of-mass time and may therefore contribute to dynamical pairing fluctuations;  see Eq.~\eqref{eq:phi_trial_omega2}.
The $\mathrm{dS}_{2}$ scattering continuum
is therefore not an auxiliary mathematical construction: it is the geometric
representation of the scale-invariant family of Eliashberg pairing modes.

More broadly, our results illustrate how a local holographic degree of freedom
can emerge from a microscopic bilocal collective field. The emergent bulk
theory is not obtained by retaining the complete formal spectrum of the
bilocal problem. Rather, the dynamics selects a low-energy spectral subspace
on which both a local gradient expansion and an invertible geometric
transformation become possible. 
The Radon transformation to anti-de Sitter space provides the most natural formulation because the continuous sector selected microscopically on the ${\rm dS}_2$ side maps onto the full normalisable spectral space of ${\rm AdS}_2$. This observation could also be useful beyond
the particular Y-SYK model considered here.

\acknowledgments

We are grateful to Andrey V. Chubukov, Ilya Esterlis, Sean Hartnoll,  Gian-Andrea Inkof, Subir Sachdev, Koenraad Schalm, and  Daniel J. Schultz for helpful discussions. We are particularly thankful for A. V. Chubukov and Shangshun Zhang for pointing out the missing ${\rm sign}$-term in the numerator of Eq.~\eqref{eq:actual_frequency_solution} in an earlier version of this manuscript.
This work was supported by 
the German Research Foundation TRR 288-422213477 ELASTO-Q-MAT, project B01 (V.C.S. and J.S.), and grant  SFI-MPS-NFS-00006741-05 from the Simons Foundation (J.S.).

\appendix
\section{Radon transform and its inverse}
\label{app:Radon}
The close connection between de Sitter and  anti--de Sitter spaces is  a consequence of the fact that one can parametrize the geodesic sub-manifolds of ${\rm AdS}_{2}$ by  points $\zeta$ in ${\rm dS}_{2}$.  This makes the integral Radon transformation~\cite{Helgason2011,Das2018,Stone2025}
\begin{equation}
\Psi\left(\zeta\right)	=	\left({\cal R} \psi\right)\left(\zeta\right)=\int_{\zeta}d\xi\sqrt{g_{\rm AdS}}\,\psi\left(\xi\right)
\end{equation}
a natural tool for relating fields on one space to fields on the other.
The integration is over the geodesics of ${\rm AdS}_{2}$, parametrized by $\zeta$, using some convenient set of coordinates $\xi$.
It assigns to a function $\psi$  over ${\rm AdS}_{2}$ a function $\Psi$ over its space of geodesics ${\rm dS}_{2}$.
The intertwinement of the Laplacians~\cite{Helgason2011,Das2018,Stone2025}
\begin{equation}
    \square_{{\rm dS}_2}{\cal R}\psi={\cal R}\left(\Box_{{\rm AdS}_2}\psi\right)
    \label{eq:Laplace_intertwinement}
\end{equation}
is a powerful property that we can employ to relate the action of a field in one space to the action in the other~\cite{Inkof2022,StangierFS2026}.
The intertwining relation applies, in particular, to the normalized eigenfunctions $\psi_{\nu}$ of $\Box_{{\rm AdS}_2}$; the Radon transform of an eigenfunction is itself an eigenfunction with the same eigenvalue. However, it is not necessarily properly normalized. Instead  one has
\begin{equation}
{\cal R}\psi_{\nu}=L_{\nu}\Psi_{\nu}\label{eq:Leg factors},
\end{equation}
where $\Psi_{\nu}$ denote the properly normalized eigenfunctions of $\Box_{{\rm dS}_2}$. The leg factors $L_{\nu}$ for the continuous part of the spectrum have been determined in Refs.~\cite{Das2018,Stone2025}. They are given as 
\begin{equation}
    L_{p}=\sqrt{\frac{2}{\pi}}\,\sqrt{\cosh\left(\pi p\right)}\left|\Gamma\left(\frac{1}{4}+\frac{ip}{2}\right)\right|^{2}.
    \label{eq:leg_factors}
\end{equation}
No leg factors for the discrete part of the spectrum exist. This is because the discrete eigenfunctions of $\Box_{{\rm dS}_2}$ can be shown to be the Radon transform of non-normalizable eigenfunctions of $\Box_{{\rm AdS}_2}$. 

The intertwinement of the Laplacians makes it possible to derive an explicit expression for the inverse of the Radon transform provided nonzero leg factors exist.  
In our case this is the subspace spanned by the continuum states. 
 To demonstrate this,  we expand $\psi$ and its Radon transform $\Psi={\cal R}\psi$
in terms of the orthonormal eigenfunctions of the two Laplacians: $\Psi\left(\zeta\right)=\sum_{\nu}A_{\nu}\Psi_{\nu}\left(\zeta\right)$ and $\psi\left(\xi\right)=\sum_{\nu}a_{\nu}\psi_{\nu}\left(\xi\right)$, respectively.
The intertwinement of the Laplacians then relates the two sets of coefficients,
\begin{equation}
    A_{\nu}=L_{\nu}a_{\nu}
    \label{eq:coeff_relation}
\end{equation}
which can be used to find for the inverse of the Radon transform
\begin{equation}
\psi\left(\xi\right)=\left({\cal R}^{-1}\Psi\right)\left(\xi\right)=\int d^{2}\zeta\sqrt{-g_{\rm dS}}\,\sum_\nu\frac{\psi_{\nu}\left(\xi\right)\Psi_{\nu}^{*}\left(\zeta\right)}{L_\nu}\Psi\left(\zeta\right).
\end{equation}
The sum over eigenstates becomes an integral when the spectrum is continuous. Hence, provided that the leg factors are non-zero, the Radon transform can be inverted. In our problem, this requires discarding the discrete part of the ${\rm dS}_2$ spectrum. The field in ${\rm AdS}_2$ can then be expressed in terms of the original field-theory degrees of freedom projected onto the continuous sector.

\section{The holographic map to ${\rm AdS}_2$}
\label{app:map}
For completeness we summarize in this appendix the holographic map from de Sitter to anti-de Sitter space as given in Refs.~\cite{Inkof2022,StangierFS2026}. We start from the Gaussian theory Eq.~\eqref{eq:act_fin} which we reproduce for convenience:
\begin{equation}
S_{{\rm pair}}=\tilde{c}_{0}N\int d^{2}\zeta\sqrt{-g_{{\rm dS}_{2}}}\Psi_{{\rm c}}^{\dagger}\left(\tilde{m}^{2}-\square_{{\rm dS}_{2}}\right)\Psi_{{\rm c}}.
\end{equation}
Here, $\Psi_{{\rm c}}$ is restricted to the continuous spectral subspace of ${\rm dS}_2$. This allows us to expand~\footnote{For simplicity, we write $\Psi_{p,\tau_{0}}$ instead of $\Psi_{\frac{1}{2}+ip,\tau_{0}}$, as would be consistent with the notation used in the main text.} 
\begin{equation}
\Psi_{{\rm c}}\left(\zeta\right)=\int_{0}^{\infty}dp\int_{-\infty}^{\infty}d\tau_{0}A_{p,\tau_{0}}\Psi_{p,\tau_{0}}\left(\zeta\right)
\end{equation}
i.e. we do not include the discrete eigenstates in the expansion.
Inserting this expansion, it follows 
\begin{equation}
S_{{\rm pair}}=\tilde{c}_{0}N\int_{0}^{\infty}dp\int_{-\infty}^{\infty}d\tau_0\, \left|A_{p,\tau_{0}}\right|^{2}\left(\tilde{m}^{2}+\frac{1}{4}+p^{2}\right).
\end{equation}
In preparation for what follows, we can equally consider a Gaussian
theory in ${\rm AdS}_{2}$ 
\begin{equation}
{\cal S}_{{\rm pair}}=c_{0}N\int d^{2}\xi\sqrt{g_{{\rm AdS}_{2}}}\psi^{\dagger}\left(m^{2}-\square_{{\rm AdS}_{2}}\right)\psi.
\label{eq:AdS_act_appendix}
\end{equation}
The spectrum of the Laplacian consists only of continuous states~\cite{Stone2025}
which allows for a similar expansion 
\begin{equation}
\psi\left(\xi\right)=\int_{0}^{\infty}dp\int_{-\infty}^{\infty}d\tau_0\,a_{p,\tau_{0}}\psi_{p,\tau_{0}}\left(\xi\right),
\end{equation}
which yields 
\begin{equation}
{\cal S}_{{\rm pair}}=c_{0}N\int_{0}^{\infty}dp\int_{-\infty}^{\infty}d\tau_0\,\left|a_{p,\tau_{0}}\right|^{2}\left(m^{2}+\frac{1}{4}+p^{2}\right).
\label{eq:AdS_act_expans}
\end{equation}
The fact that the Radon transform is invertible is reflected in the
leg factors of Eq.~\eqref{eq:leg_factors}, which, according to Eq.~\eqref{eq:coeff_relation}, relate the expansion
coefficients: $A_{p,\tau_{0}}=L_{p}a_{p,\tau_{0}}$.
As we are considering the limit of small $p$ we can expand Eq.~\eqref{eq:leg_factors}
\begin{equation}
    L_{p}\approx\sqrt{\frac{2}{\pi}}\Gamma\left(\frac{1}{4}\right)^{2}\left(1-2Gp^{2}\right)
\end{equation}
with Catalan constant $G\approx0.91597$. Expressing $S_{{\rm pair}}$
in terms of the expansion coefficients $a_{p,\tau_{0}}$ of ${\rm AdS}_{2}$ and  keeping only terms of order $p^{2}$,  we obtain
\begin{equation}
S_{{\rm pair}}=c_{0}N\int dpd\tau_{0}\left|a_{p,\tau_{0}}\right|^{2}\left(m^{2}+\frac{1}{4}+p^{2}\right),
\end{equation}
where $c_{0}=\sqrt{\frac{2}{\pi}}\Gamma\left(\frac{1}{4}\right)^{2}\left(\tilde{c}_{0}-G\left(1-k_{0}\right)\right)$
and 
\begin{equation}
m^{2}=\frac{\tilde{m}^{2}+\frac{1}{4}}{1-2G\left(\tilde{m}^{2}+\frac{1}{4}\right)}-\frac{1}{4}.
\end{equation}
With Eq.~\eqref{eq:AdS_act_expans} we can  identify this expression with the action Eq.~\eqref{eq:AdS_act_appendix} in ${\rm AdS}_{2}$. The masses $m$ and ${\tilde m}$ agree near the Breitenlohner-Freedman bound. This demonstrates that the continuous part of the de Sitter action
can be expressed as a theory in anti-de Sitter space.

\begingroup
\section{Kernel eigenfunctions and their conformal degeneracy}
\label{app:eigenfct_proof}
In this appendix, we show that $F_{h,\tau_{0}}\left(\tau_{1},\tau_{2}\right)$
defined in Eq.~\eqref{eq:eigenf_gen} is an eigenfunction of the kernel $K_{F}$, with an
eigenvalue independent of $\tau_{0}$. In view of the similarity
transformations relating the different representations, it suffices
to analyse the kernel $K_{\Phi}$ and the functions
\begin{equation}
\Phi_{h,\tau_{0}}\left(\tau_{1},\tau_{2}\right)=\frac{{\rm sign}\left(\tau_{1}-\tau_{0}\right){\rm sign}\left(\tau_{2}-\tau_{0}\right)|\tau_{0}|^{2h}}{|\tau_{1}-\tau_{0}|^{h}|\tau_{2}-\tau_{0}|^{h}|\tau_{1}-\tau_{2}|^{2-h-2\Delta}}.\label{eq:eigenfunctions}
\end{equation}
The integrals that occur upon applying $K_\Phi$  are absolutely convergent for
$\frac{1}{4}<\Delta<\frac{1}{2}$ and
$1-2\Delta<{\rm Re}\left(h\right)<2\Delta$. Outside this strip,
in particular for the discrete part of the spectrum,
$k\left(h\right)$ must be obtained by analytic continuation,
and the corresponding integrals require regularisation.
The eigenfunctions Eq.~\eqref{eq:eigenfunctions} are not normalized. The natural scalar product is
\begin{equation}
\left\langle \Phi_{1}\mid\Phi_{2}\right\rangle =\int d\tau_{1}\int d\tau_{2}\frac{\Phi_{1}^{*}\left(\tau_{1},\tau_{2}\right)\Phi_{2}\left(\tau_{1},\tau_{2}\right)}{\left|\tau_{1}-\tau_{2}\right|^{4\Delta-2}},
\end{equation}
with respect to which $K_{\Phi}$ is Hermitian. For continuum states
with $h=\frac{1}{2}+ip$ and $h'=\frac{1}{2}+ip'$, the overlap takes
the form
\begin{equation}
\left\langle \Phi_{h,\tau_{0}}\mid\Phi_{h',\tau'_{0}}\right\rangle =\frac{4\pi^{2}|\tau_{0}|^{{\rm Re} (2h)}}{p}\coth\left(\pi p\right)\delta\left(p-p'\right)\delta\left(\tau_{0}-\tau'_{0}\right).
\end{equation}
We do not impose a specific normalisation in what follows.

In the limit $\tau_{0}\rightarrow\infty$, Eq.~(\ref{eq:eigenfunctions})
reduces to
\begin{equation}
\Phi_{h,\infty}\left(\tau_{1},\tau_{2}\right)=\frac{1}{|\tau_{1}-\tau_{2}|^{2-h-2\Delta}}.
\end{equation}
Substitution into the eigenvalue equation for $K_{\Phi}$ confirms,
after some algebra, that $\Phi_{h,\infty}$ is an eigenfunction
and yields the eigenvalue $k\left(h\right)$ given in Eq.~\eqref{eq:eigenvals_K_fin}.
We now show that $\Phi_{h,\tau_{0}}$ in Eq.~(\ref{eq:eigenfunctions})
is an eigenfunction for arbitrary $\tau_{0}$, with the same
eigenvalue $k\left(h\right)$. To this end, we introduce the
M{\"o}bius transformation
\begin{equation}
T\left(\tau\right)=\frac{\tau}{1-\tau/\tau_{0}}.
\end{equation}
It transforms  $\tau=\tau_0$ to $T(\tau_0)\rightarrow \pm \infty $.
We write for $i=1,2$: $d_{i}=1-\tau_{i}/\tau_{0}$,
$s_{i}={\rm sign}\left(d_{i}\right)$, and
$T_{i}=T\left(\tau_{i}\right)$, which  implies
\begin{equation}
T_{i}-T_{j}=\frac{\tau_{i}}{d_{i}}-\frac{\tau_{j}}{d_{j}}=\frac{\tau_{i}-\tau_{j}}{d_{i}d_{j}}.
\end{equation}
We further define the transformation operator $U_{\tau_{0}}$,  associated with this M{\"o}bius transformation, by:
\begin{equation}
\left(U_{\tau_{0}}\Phi\right)\left(\tau_{1},\tau_{2}\right)=\frac{s_{1}s_{2}}{\left|d_{1}d_{2}\right|^{2-2\Delta}}\Phi\left(T_{1},T_{2}\right).\label{eq:conf_mapdef}
\end{equation}
It allows to generate the function at finite $\tau_{0}$   from $\Phi_{h,\infty}$ through
\begin{equation}
\left(U_{\tau_{0}}\Phi_{h,\infty}\right)\left(\tau_{1},\tau_{2}\right) =  \Phi_{h,\tau_{0}}\left(\tau_{1},\tau_{2}\right).\label{eq:tao_0transd}
\end{equation}

Next we we show that the operator $U_{\tau_{0}}$ commutes with $K_{\Phi}$.
The power-law factors in the kernel transform as
\begin{eqnarray}
\frac{{\rm sign}\left(T_{i}-T_{j}\right)}{|T_{i}-T_{j}|^{2\Delta}} & = & s_{i}s_{j}|d_{i}d_{j}|^{2\Delta}\frac{{\rm sign}\left(\tau_{i}-\tau_{j}\right)}{|\tau_{i}-\tau_{j}|^{2\Delta}},\nonumber \\
\frac{1}{|T_{1}-T_{2}|^{2-4\Delta}} & = & |d_{1}d_{2}|^{2-4\Delta}\frac{1}{|\tau_{1}-\tau_{2}|^{2-4\Delta}}.
\end{eqnarray}
Combining these factors yields
\begin{equation}
K_{\Phi}\left(T_{1},T_{2};T_{3},T_{4}\right)=s_{1}s_{2}s_{3}s_{4}\left|d_{1}d_{2}\right|^{2-2\Delta}\left|d_{3}d_{4}\right|^{2\Delta}K_{\Phi}\left(\tau_{1},\tau_{2};\tau_{3},\tau_{4}\right).
\end{equation}
Using the transformed integration measure,
$dT_{i}=d_{i}^{-2}d\tau_{i}$, we obtain
\begin{eqnarray}
\left(U_{\tau_{0}}K_{\Phi}\Phi\right)\left(\tau_{1},\tau_{2}\right) & = & \frac{s_{1}s_{2}}{\left|d_{1}d_{2}\right|^{2-2\Delta}}\int dT_{3}dT_{4}K_{\Phi}\left(T_{1},T_{2};T_{3},T_{4}\right)\Phi\left(T_{3},T_{4}\right)\nonumber \\
 & = & \int dT_{3}dT_{4}\left|d_{3}d_{4}\right|^{2}K_{\Phi}\left(\tau_{1},\tau_{2};\tau_{3},\tau_{4}\right)\left(U_{\tau_{0}}\Phi\right)\left(\tau_{3},\tau_{4}\right)\nonumber \\
 & = & \left(K_{\Phi}U_{\tau_{0}}\Phi\right)\left(\tau_{1},\tau_{2}\right).
\end{eqnarray}
Thus, $\left[U_{\tau_{0}},K_{\Phi}\right]=0$, and consequently
\begin{equation}
K_{\Phi}\Phi_{h,\tau_{0}}=K_{\Phi}U_{\tau_{0}}\Phi_{h,\infty}=U_{\tau_{0}}K_{\Phi}\Phi_{h,\infty}=k\left(h\right)\Phi_{h,\tau_{0}}.
\end{equation}
Hence, the functions $\Phi_{h,\tau_{0}}$ are eigenfunctions
of $K_{\Phi}$ with eigenvalue $k\left(h\right)$ independent
of $\tau_{0}$, establishing the claimed degeneracy.
\endgroup



\bibliographystyle{JHEP}
 \bibliography{refs}


\end{document}